# Fourier Analysis of an Expanded Gravity Model for Spatio-Temporal Interactions

Yanguang Chen[1], Fahui Wang[2]

(1. Department of Geography, Peking University, Beijing 100871, PRC, E-mail: chenyg@pku.edu.cn; 2.Department of Geography, Louisiana State University, Baton Rouge, LA 70803, USA, E-mail: fwang@lsu.edu.)

**Abstract:** Fourier analysis and cross-correlation function are successfully applied to improving the conventional gravity model of interaction between cities by introducing a time variable to the attraction measures (e.g., city sizes). The traditional model assumes spatial interaction as instantaneous, while the new model considers the interaction as a temporal process and measures it as an aggregation over a period of time. By doing so, the new model not only is more theoretically sound, but also enables us to integrate the analysis of temporal process into spatial interaction modeling. Based on cross-correlation function, the developed model is calibrated by Fourier analysis techniques, and the computation process is demonstrated in four steps. The paper uses a simple case study to illustrate the approach to modeling the interurban interaction, and highlight the relationship between the new model and the conventional gravity model.

## 1 Introduction

The gravity model is an equation of the interaction between two population centers based on Newton's Law of Universal Gravitation. Two centers in a geographical region attract one another in proportion to the product of their "masses" and inversely as the square distance between them. The original equation has been changed to accommodate the special needs of geographical systems. So the impedance function is not confined to the reciprocal of the distance squared, and the application of the model is not limited to urban geography. In fact, the gravity model may be applied to fields of influence of settlements, migration of population, trade, traffic flows, and telephone calls, etc (Erlander, 1980; Fotheringham and O'Kelly, 1989; Haynes and Fotheringham, 1984; Lierop, 1986; Roy, 2004; Sen and Smith, 1995). However, our attention will focus only on influence of cities or towns.

The gravity model was first proposed because of its ability to replicate observed urban flow patterns, and thus was purely empirical (Carrothers, 1956). Theoretical justifications for the gravity model were later proposed by (1) Wilson (1970, 2000) using the entropy-maximizing principle (also see Tomlin and Tomlin, 1968; Curry, 1972), and (2) Niedercorn and Bechdolt (1969) based on individual utility-maximizing behavior (also see Golob and Beckmann, 1971; Allen, 1972; Colwell, 1982). In fact, one of the major achievements in the era of "quantitative revolution" in geography was the proof that the micro models of spatial allocation of individual resources based on utility-maximizing principle were consistent with the macro models of spatial interaction based on the entropy-maximizing principle (Batty, 2000). As applications of the gravity models grow, there have been persistent inquiries for theoretical foundations of the gravity models. Among recent examples, Bavaud (2002) used the quasi-symmetric property in Markov chains theory, Bradley-Teryy-Luce decision theory and others to unify various traditions in gravity modeling; and Evenett and Keller (2002) found the explanations from two important trade theories, namely the Heckscher-Ohlin theory and the increasing returns theory, and defined the conditions for the empirical success of the gravity equations. Clearly, the gravity model has solid foundations on well-known theories, and its wide applicability in predicting socioeconomic interactions is no surprise to geographers and other social scientists.

However, despite varied theoretical discussions and its application to various fields, the conventional gravity model has a fatal weakness owing to lack of temporal dimension. This weakness presents an obstacle to the development of spatial interaction theory. In fact, the interaction of human phenomena (e.g. cities) is different from that of physical phenomena (e.g. celestial bodies). For human systems, there exists a delay not to be ignored between action and reaction which has been ignored for a long time past. In this paper, we will make the conventional model into a new expression by introducing the time lag of spatial interaction to gravity modeling.

Thus the product of attraction measures become a cross-correlation function, and then Fourier analysis is an indispensable tool. The expanded gravity model measures the aggregated interaction over time instead of interaction at a certain time point, and attraction measures become a function of time lag. The new model is not only more theoretically sound, but also enables us to examine the temporal process of spatial interaction.

Fourier analysis is of significance for solving the difficult problems of geographical research (Chen and Liu, 2006; Chen and Zhou, 2008). Once we apply the method to modeling the spatio-temporal interaction of cities, we will find a surprising but interesting and satisfying result. The remaining part of this paper is structured as follows. Section 2 introduces the expanded gravity model based on the cross-correlation function, and discusses the model's solution based on Fourier analysis. A new concept based on power spectrum, *gravity spectrum*, is proposed to characterize spatial interaction of cities. Section 3 explains the model's calibration by discrete Fourier analysis and implementation in common software such as the Microsoft Excel. Section 4 uses a simple case study to illustrate the modeling, analyze the temporal process of spatial interaction, and highlight the relationship between the new model and the conventional gravity model. Finally, the paper is concluded with a brief summary.

## 2 The Expanded Gravity Model Based on Cross-Correlation

The general gravity model of interaction between city *i* and city *j* can be written as

$$I_{ij} = KQ_iQ_j f(r),  \qquad (1)$$

where the interaction $I_{ij}$ between two cities is positively related to their sizes $Q_i$, $Q_j$, and inversely related to the distance *r* between them, and *K* is a scalar constant. The distance impedance *f(r)* is commonly measured in three forms such as (Kwan, 1998): (1) an inverse power function $f(r) = r^{-b}$, (2) a negative exponential function $f(r) = e^{-br}$, or (3) a Gaussian function $f(r) = e^{-br^2}$, where *b* is a constant. Choice of impedance/resistance function has potential ramifications on model estimation (see Fotheringham and O'Kelly, 1989). Without losing generality, this paper uses the most common form $f(r) = r^{-b}$, for the distance impedance function.

In the conventional gravity model, equation (1), interaction between cities (e.g., in terms of traffic, trade, communication or migration) is assumed instantaneous, and attraction measures of the cities (e.g., in terms of sizes of population or economy) are static. In this research, we argue that the development of spatial interaction in socioeconomic context takes time, and there is a time lag in the attraction measures of cities. The lag may be short in some cases but long in others. Now considering each of the city sizes $Q_i$ and $Q_j$ as a variable of time $t$: $Q_i = f_i(t)$ and $Q_j = f_j(t)$, and introducing a time-lag parameter $\tau$ to capture the time gap between them for the interaction to take place, the gravity model based on the inverse power function in equation (1) is rewritten as

$$I_{ij}(t) = K f_i(t) f_j(t+\tau) r^{-b}. \qquad (2)$$

For theoretical considerations, it is assumed that both $Q_i = f_i(t)$ and $Q_j = f_j(t)$ are bounded. This is a reasonable assumption as city sizes have limits, particularly in finite time.

In contrast to the conventional gravity model in equation (1), the new model in equation (2) has two new elements: First, the interaction $I_{ij}$ (e.g., commodity or passenger flow volume) becomes *directional*, and $I_{ij}$ is not necessarily equal to $I_{ji}$ as is commonly observed in the real world. For instance, Colwell (1982) argues that residents in a lower-level (smaller) need to make more trips to a higher-level (larger) city than the other way around as suggested by the central place theory. However, sometimes the opposite is true in the real world, especially for commodity flows. Second, as shown in Figure 1, there is a *time lag* for spatial interaction to reach the destination city $j$ (at time $t+\tau$) from the origin city $i$ (at time $t$). The interaction $I_{ij}$ is from city $i$ at time $t$ to city $j$ at time $t+\tau$, and similarly, the interaction $I_{ji}$ would be from city $j$ at time $t$ to city $i$ at time $t+\tau$.

By introducing a time variable to the conventional gravity model, the spatial interaction between cities becomes a continuous temporal process, and the expanded gravity model captures a *complex spatio-temporal system*. Equation (2) measures the *instantaneous* interaction between cities $i$ and $j$ at a particular time. City sizes change over time, and interactions between them in terms of traffic, information or financial transaction flows are usually measured for a period of time. Aggregating the instantaneous interaction over time generates the interaction between the two cities during a period of time. Suppose that the life-span of all cities are finite, i.e., if $t \to \infty$, then $Q_i$, $Q_j \to 0$. In formula, we have a new interaction model as follows

$$G_{ij}(\tau) = \int_{-\infty}^{\infty} I_{ij}(t)dt = Kr^{-b}\int_{-\infty}^{\infty} f_i(t)f_j(t+\tau)dt = KR_{ij}(\tau)r^{-b}, \qquad (3)$$

where

$$R_{ij}(\tau) = \int_{-\infty}^{\infty} f_i(t)f_j(t+\tau)dt \qquad (4)$$

defines a cross-correlation function between cities $i$ and $j$ with a time lag $\tau$ given that the data are standardized (Bracewell, 2000; Boggess and Narcowich, 2002).

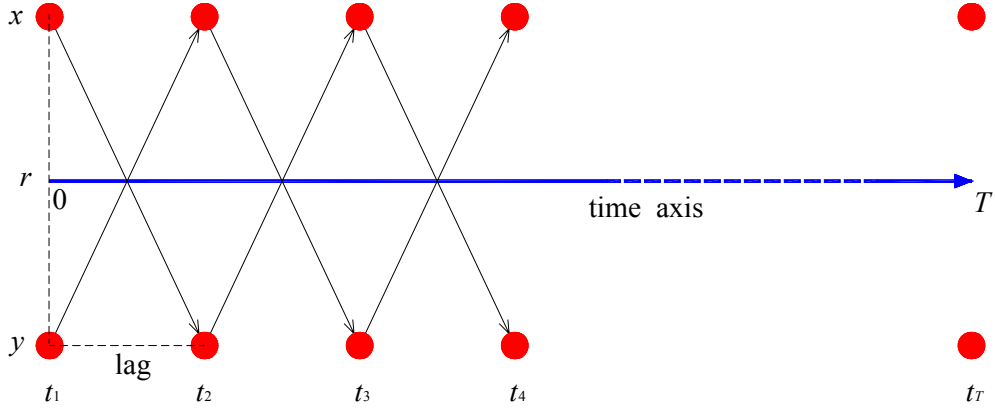

Figure 1. Spatial interactions between two cities with a time lag

Using the power theorem (See e.g. Bracewell, 2000)

$$\int_{-\infty}^{\infty} f_i(t)f_j(t)dt = \frac{1}{2\pi}\int_{-\infty}^{\infty} \overline{F_i(\omega)}F_j(\omega)d\omega \qquad (5)$$

and the Fourier transform of a translation (See e.g. Boggess and Narcowich, 2002)

$$\mathscr{F}[f(t+\tau)] = F(\omega)e^{i\omega\tau}, \qquad (6)$$

we obtain the relation between the *correlation function* $R(\tau)$ and *energy spectra density* $S(\omega)$ such as

$$R_{ij}(\tau) = \int_{-\infty}^{\infty} f_i(t)f_j(t+\tau)dt = \frac{1}{2\pi}\int_{-\infty}^{\infty} \overline{F_i(\omega)}F_j(\omega)e^{i\omega\tau}d\omega = \mathscr{F}^{-1}[S_{ij}(\omega)], \qquad (7)$$

where $\mathscr{F}$ is the operator of Fourier transform (FT) while $\mathscr{F}^{-1}$ the operator of inverse Fourier transform (IFT), and

$$S_{ij}(\omega) = \overline{F_i(\omega)}F_j(\omega) \qquad (8)$$

is the cross energy spectrum of a pair of series $f_i(t)$ and $f_j(t)$ (Bracewell, 2000; Boggess and Narcowich, 2002), $F(\omega)=\mathscr{F}[f(t)]$ denotes the Fourier transform of $f(t)$, and $\omega=2\pi f$ is angular frequency ($-\infty<\omega<\infty$, where $f$ is linear frequency). In other words, the cross-correlation function $R(\tau)$ can be computed through the IFT of cross energy spectrum $S_{ij}(\omega)$. This is one of the keys for us to improve the gravity model of cities using the idea from Fourier analysis and correlation function.

Therefore, equation (3) can be rewritten as

$$G_{ij}(\tau) = KR_{ij}(\tau)r^{-b} = K\mathscr{F}^{1}[S(\omega)]r^{-b}. \tag{9}$$

Equation (9) defines the *spatial interaction* (or *gravity*) *spectrum* that varies with the time lag $\tau$. In other words, it captures how the interaction between cities varies with different lengths of time lag. The correlation function, as specified in equation (4), is the basic component in the expanded gravity model. The new model in the current form focuses on the distance-based correlation between city sizes *over time* whereas the conventional gravity model emphasizes the effect of distance decay *over space*. As it is known in physics, gravitational force multiplied by action distance is work, and the work per unit time is power. Naturally, the power spectral analysis techniques in physics can be applied to analysis of the expanded gravity model in socioeconomic context, as demonstrated in the next section.

## 3 Model's Calibration

### 3.1 Discrete Fourier Analysis

The expanded gravity model in equation (3) is a continuous function with infinite time span. It is only possible to solve the equation through the Fourier transform in equation (9) if the function is relatively uncomplicated (Weaver, 1983). In the real world, the function is represented by a sequence of numbers. If the observations are made over time, it is a time series, and discrete Fourier analysis is used for analysis of time series (Bloomfield, 2000). The following illustrates the discrete Fourier Transform (DFT) and its relationship to the commonly-used correlation analysis.

The correlation coefficient between data series $x$ and $y$ can be written as

$$\rho_{xy}(k) = \frac{C_{xy}(k)}{\sigma_x \sigma_y}, \tag{10}$$

where $\sigma_x$ and $\sigma_y$ are the standard deviations of $x_t$ and $y_t$ respectively, $C_{xy}(k)$ is the correlation function or in fact covariance defined as

$$C_{xy}(k) = \frac{1}{N-k} \sum_{t=1}^{N-k} (x_t - \mu_x)(y_{t+k} - \mu_y), \tag{11}$$

where $\mu_x$ and $\mu_y$ are the means of $x_t$ and $y_t$ respectively. Note that $\rho_{xy}(k)$ is equivalent to $C_{xy}(k)$ with a constant multiplier. The following analysis focuses on $C_{xy}(k)$. Equation (10) is the discrete version of equation (7). When the variables are standardized with $\mu_x=\mu_y=0$, and $\sigma_x=\sigma_y=1$, clearly $\rho_{xy}(k)$ in a discrete form corresponds to a continuous function $R_{ij}(\tau)$. The parameter $k$ ($=0, \pm1, \pm2, \cdots$) in equation (10) is time lag in the discrete series.

One may attempt to calibrate the expanded gravity model directly based on the correlation function such as

$$G_{xy}(k) = K r^{-b} C_{xy}(k). \tag{12}$$

However, the direct computation of correlation quantity at various time lags is cumbersome, and determining the maximum time lag is problematic, the results depending often on researchers' experiences. As illustrated previously in continuous functions, the analysis of $R_{ij}(\tau)$ in the time domain can be converted to the analysis of $S_{ij}(\omega)$ in the frequency domain using Fourier transforms. The same strategy is adopted here in discrete forms.

The discrete Fourier transform (DFT) of $C_{xy}(k)$ is written as

$$P_{xy}(f) = \sum_{k=0}^{N-1} C_{xy}(k) e^{-i2\pi \frac{m}{N} k} = \sum_{k} C_{xy}(k) e^{-i2\pi f k}, \tag{13}$$

where $P(f)$ is the power spectral density of $C_{xy}(k)$, and the linear frequency is measured as $f=m/N$, where $m=0, 1, 2, \cdots, N-1$, and $N$ is the length of sample path. Similar to the continuous function $R(\tau)$ in equation (7), the calibration of $P(f)$ is obtained through discrete power spectrum analysis such as

$$P_{xy}(\omega) = \frac{1}{2\pi N} \overline{F_x(\omega)} F_y(\omega), \tag{14}$$

In practice, the angular frequency $\omega$ is replaced by the linear frequency $f$ for convenience, namely $\omega = 2\pi f = 2\pi n/N$. Therefore, $P(f)$ is rewritten in the following form

$$P(f) = \frac{1}{N} \overline{F_x(f)} F_y(f), \tag{15}$$

As $P(f)$ is the DFT of $C_{xy}(k)$, $C_{xy}(k)$ is the inverse discrete Fourier transform (IDFT) of $P(f)$, that is

$$C_{xy}(k) = \frac{1}{N} \sum_{n=0}^{N-1} P_{xy}(f) e^{i2\pi \frac{m}{N} k} = \frac{1}{N} \sum_f P_{xy}(f) e^{i2\pi f k} = \text{IDFT}[P_{xy}(f)]. \tag{16}$$

That implies $C_{xy}(k) \propto R_{xy}(\tau)/N$. Applying a multiplier $Kr^{-b}$ to completing the calibration of the expanded gravity model, we have

$$G_{xy}(k) = Kr^{-b} \text{IDFT}[P_{xy}(f)]. \tag{17}$$

This equation will give a pair of discrete gravity spectra of urban interaction.

**3.2 Implementation in Excel**

In applications, the fast Fourier transform (FFT) algorithm is employed to perform the Fourier analysis. Various mathematical software packages may be used to implement the FFT. The following uses the Fourier Analysis Toolpak available in MS Excel (Bloch, 2002) to explain the model's implementation step by step. In fact, if a computation can be conducted in MS Excel to find the solution to a problem, it will be able to be completed more easily in some mathematical software such as Mathcad, Matlab, etc. by designing programs. In Excel, one can access the toolpak by choosing "Tools", then "Data Analysis", and then "Fourier Analysis". Note that "inverse Fourier transform" is also available by choosing "inverse" in the same dialogue box. The application of the new method to interaction of cities can be summarized in four steps:

(1) Performing the FFT on $f_x(t)$ and $f_y(t)$ yields two complex data series $F_x(f)$ and $F_y(f)$. Note that we use the linear frequency measure $f$ instead of the angular frequency measure $\omega$ here. The outputs (complex data series) are shown as non-numerical texts in Excel.

(2) Compute the complex conjugate of $F_x(f)$, i.e., $\overline{F_x(f)}$, and then calculate the product of two complex data series, i.e., $\overline{F_x(f)}F_y(f)$. Dividing the result by the number of data points, $N$, yields the *cross power spectrum*, $P(f)$, as in equation (15).

(3) Performing the inverse fast Fourier transform (IFFT) on $P(f)$ yields $C_{xy}(k)$. This implements the calibration of equation (16).

(4) Applying a multiplier $Kr^{-b}$ to the cross-correlation function yields the interaction from city $x$ to city $y$, $G_{xy}(k)$, as indicated by equation (17).

Similarly, one may follow steps (2)-(4) by reversing the orders of $x$ and $y$ to obtain the interaction from city $y$ to city $x$ such as $G_{yx}(k) = Kr^{-b}C_{yx}(k)$. The above computation process yields two spectra characterizing the spatio-temporal interaction process between two cities.

## 4 An Empirical Example

This section uses an empirical example to illustrate how to calibrate the expanded gravity model by the discrete Fourier analysis based on FFT, and discuss the implications of the model. The case study is selected both for the convenience of data availability and for clarity of demonstrating the computation process. Therefore, only two Chinese cities, Beijing and Tianjin, are taken into consideration. In this case, we set the distance variable in the gravity model as a constant, and focus on the new element of time variable. This is a valid approach in many applications in time-series analysis. For instance, in studying the stock market correlations over time, Flavin *et al.* (2002) used the gravity model with a fixed distance and focused on the interaction over time. Apparently, it is not difficult for us to employ the model in the case that the distance is a variable.

### 4.1 Spatial Interaction Spectra between Beijing and Tianjin 1949-2004

Beijing and Tianjin are two large cities in northern China that are only 137 kilometers apart from each other. They can be regarded as a "double star" in the constellation of Chinese cities despite the fact that Tianjin was originally a satellite town of Beijing hundred years ago (Chen and Zhou, 2003). The strong social and economic interactions between them have been well documented in the history. The Beijing-Tianjin Railway, designed by the legendary Chinese railroad engineer Zhan Tianyou, was the first railroad constructed in China as early as in 1902. The commercial flight between Beijing and Tianjin was also the first in China in 1920 (Jin *et al*, 2004). Today traffics on the highways and railway connecting the two cities are among the busiest in China. This research uses "non-agricultural population", the most reliable measure for urban population sizes (particularly for historical data) in China (Zhou, 1995), to represent their sizes $Q_B$ and $Q_T$. Data of non-agricultural population in Beijing and Tianjin from 1949 to 2004 are extracted from statistical reports compiled by the National Bureau of Statistics of China (See the Appendix 1 for the data).

The process for calibrating the expanded gravity model has been illustrated in section 3.2. The following follows the same process for further clarification.

Step 1 performs the FFT on the population series of Beijing and Tianjin respectively. The symmetrical rule in the FFT's recursive algorithm requires the length of time series to be an integer power of 2, i.e., $N=2^n$ ($n=1, 2, 3, \cdots,$ ). However, there are 56 data points in each series. A process called "zero-padding" is used to bring the number up to the next power of 2 (Bloch, 2002). In this case, adding 8 zeros at the end of the 56-year series brings the number to 64 (i.e., $N=2^6$). Adding more zeros to bring the number to 128 would artificially decrease the variance without adding any more information; and truncating the data series to 32 implies that not all data points would be utilized After the zero-padding process, the FFT is performed on the data series of Beijing (denoted by $f_B(t)$) and that of Tianjin (denoted by $f_T(t)$), i.e.,

$$F_B(f) = \text{FFT}[f_B(t)], F_T(f) = \text{FFT}[f_T(t)].$$

The resulting complex data series are written as $F_B(f) = a_B + ib_B$ and $F_T(f) = a_T + ib_T$.

Step 2 computes the cross spectral density such as

$$P_{BT}(f) = \frac{1}{T}\overline{F_B(f)}F_T(f) = \frac{1}{64}[(a_Ba_T + b_Bb_T) + (a_Bb_T - b_Ba_T)i],$$

$$P_{TB}(f) = \frac{1}{T}\overline{F_T(f)}F_B(f) = \frac{1}{64}[(a_Ba_T + b_Bb_T) - (a_Bb_T - b_Ba_T)i].$$

Evidently, $P_{TB}(f)$ is the conjugate complex of $P_{BT}(f)$, i.e.,

$$\overline{P_{BT}(f)} = P_{TB}(f).$$

In other words, one may obtain $P_{TB}(f)$ once $P_{BT}(f)$ is calculated.

As illustrated in step (2) in section 3.2, $P_{BT}(f)$ is obtained by (i) using the function IMCONJUGATE to calibrate $\overline{F_B(f)}$, (ii) using the function IMPRODUCT to compute $\overline{F_B(f)} \times F_T(f)$ and, (iii) using the function IMPRODUCT again to compute $\overline{F_B(f)} F_T(f) \times (1/64)$. Finally, $P_{TB}(f)$ can be gotten by using the function IMCONJUGATE over again.

Step 3 calculates the cross-correlation functions by performing the IFFT on the cross spectral density, i.e.,

$$C_{BT}(k) = \text{IFFT}[P_{BT}(f)],$$

$$C_{TB}(k) = \text{IFFT}[P_{TB}(f)] = \text{IFFT}[\overline{P_{BT}(f)}].$$

Theoretically, the result from IFFT should be a data series of real numbers expressed with complex numbers. In other words, the output is a special series of complex numbers, the imaginary parts of which are expected to be 0. However, due to computation errors, the imaginary parts may not be, but very close to, 0. Thus, it is necessarily to use the function IMREAL to obtain the real parts of the result.

In the final step, setting the scalar at $K=1$ for simplicity, taking the average distance by railway and highways between Beijing and Tianjin as $r=137$ km, and assuming the distance friction coefficient $b=2$ as usual, the multiplier is obtained such as $Kr^{-b} = 1/137^2$. Therefore, the gravity spectra of spatial interaction between Beijing and Tianjin for 1949-2004 can be given by

$$G_{BT}(k) = C_{BT}(k) \times 137^{-2},$$

$$G_{TB}(k) = C_{TB}(k) \times 137^{-2}.$$

As $F_{BT}(k) \neq F_{TB}(k)$ in general, Figure 2 shows two spectra for spatial interactions between Beijing and Tianjin given various time lags. Based on the symmetric property of the spectra along the horizontal axis (discrete time lag $k$), Figure 2 shows only the left half of the gravity spectra ($k = 1, 2, \cdots, 32$).

The aforementioned procedure of calculation can be carried out easily with other mathematical software, e.g., Matlab, Mathcad, by designing programs.

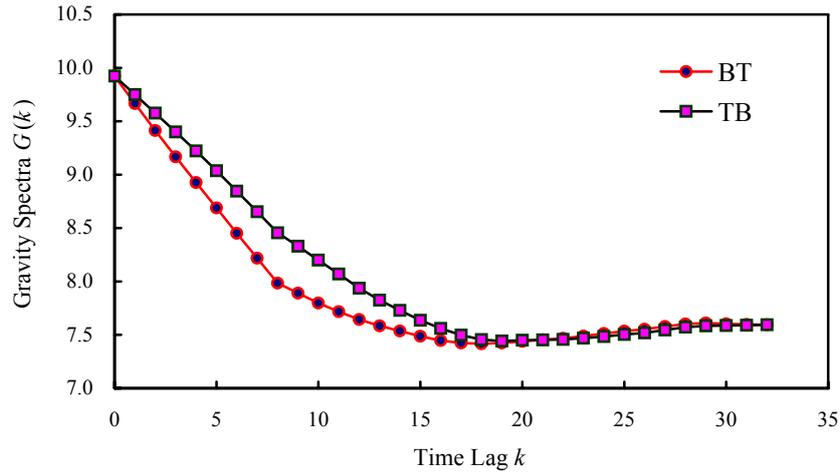

**Figure 2. Spatial interaction spectra between Beijing and Tianjin: 1949-2004**

(**Note**: BT indicates the force from Beijing to Tianjin, and TB from Tianjin to Beijing.)

Two important observations can be made from Figure 2. First, *there are time lags in spatial interactions between the two cities*. Figure 2 shows that the spectra decline gradually and level off after 15-16 years. In other words, the interactions get weaker with a longer time lag and last up to 15-16 years and tail off afterwards. Second, *the interactions are asymmetric in directions*. Figure 2 shows that the spectrum for the Beijing-Tianjin interaction is below the Tianjin-Beijing spectrum. Take the commodity flow as an indication of interaction. Given the larger city size of Beijing than Tianjin, this indicates that the flow volume is likely to be higher in the direction from a larger city to a smaller city (indicating the drawing force of a smaller on a larger city) than the reverse direction. This is consistent with the actual freight flows between Beijing and Tianjin 1985-2000 (see table 1).

**Table 1. Freight flows between Beijing and Tianjin: 1985-2000 (in 10,000 tons)**

| Year | Beijing → Tianjin | Tianjin → Beijing |
| --- | --- | --- |
| 1985 | 384 | 221 |
| 1990 | 347 | 205 |
| 1995 | 385 | 203 |
| 2000 | 518 | 318 |

**Data sources**: *China Transportation Annual Reports* in 1986, 1991, 1996, and 2001 published by China Transportation Annual Reports Press.

## 4.2 Relationship to the Conventional Model and Comparison

First of all, the expanded gravity model is regressed to the traditional model when the time lag $k$ = 0. If $k$ = 0, then $C_{BT}(0) = C_{TB}(0) = 186247.826$, and $F_{BT}(0) = F_{TB}(0) = C(0) \times 137^{-2} = 9.923$, which is the intercept as shown in Figure 2. The conventional gravity model in equation (1) only defines the interaction at a given time. For instance, given the same parameters, the interaction between Beijing and Tianjin in 2004 is $I_{BT} = KP_B P_T r^{-b} = 1 \times 854.70 \times 556.17 \times 137^{-2} = 25.327$. Repeat calibration of the traditional model for each year, and calculate the average value of spatial interactions $I_{BT}$ between the two cities for 1949-2004. The result is 9.923, exactly equal to the spectrum density by the expanded model at $k$ = 0.

This property can be proven as follows. From equation (7), we have

$$C_{ij}(\tau) = \frac{1}{T} R_{ij}(\tau) = \frac{1}{T} \int_0^T f_i(t) f_j(t+\tau) \mathrm{d}t = \mathscr{F}^{-1}[P_{ij}(\omega)]. \qquad (18)$$

For the discrete series, given time lag $\tau = 0$, equation (12) becomes

$$G_{xy}(0) = Kr^{-b} \frac{1}{N} \sum_{t=1}^{N} f_x(t) f_y(t) = \frac{1}{N} \sum_{t=1}^{N} I_{xy}(t). \qquad (19)$$

That is to say, when the time lag is 0, the gravity calibrated from the expanded model is the average of gravity values obtained from the traditional model over a study period. Owing to the Fourier transform is based on FFT, the divisor of averaging the set of interaction values $I_{xy}(t)$ will be 64 instead of 56 in our example.

To further enhance the understanding of the new model, we use the traditional model to calculate $I_{xy}$ but match the origin city size $x$ with the destination city size $y$ with a time lag $k$, and then compute the average value of $I_{xy}$ across the time span. The results with different time lags $k$, presented in Table 2, are indeed equal to the spectrum density obtained by the new model. For example, for $k$ = 3, we use the conventional gravity model to calculate $I_{BT}(1)$ based on Beijing's population in 1949 and Tianjin's population in 1952, $I_{BT}(2)$ using Beijing's population in 1950 and Tianjin's population in 1953, **…**; and then compute the average $\bar{I}_{BT} = [I_{BT}(1) + I_{BT}(2) + \ldots + I_{BT}(53)]/T = 9.167$, which is equal to the spectrum density $G_{BT}(3) = 9.167$ based on the new model. On the other side, the gravity value from Tianjin to Beijing is $\bar{I}_{TB} = G_{TB}(3) = 9.400$.
By the way, these averages are different from those in the common sense. There is an analogy between the way we determine the means here and the way cross-correlation coefficients are calculated in time series analysis. That is, we divide the sums by *T*, even though only *(T-τ)* terms appear in the sum (For further information, see Box *et al*, 1994; Diebold, 2001).

Table 2. Spatial interaction by the conventional and the expanded gravity models

| Time lag (k) | Gravity spectrum density by the expanded model | | Average spatial interaction by the conventional model | |
|---|---|---|---|---|
| | Beijing-Tianjin | Tianjin-Beijing | Beijing in *t*-th Year- Tianjin in (*t+k*)-th Year | Tianjin in *t*-th Year- Beijing in (*t+k*)-th Year |
| k=0 | 9.923 | 9.923 | 9.923 | 9.923 |
| k=1 | 9.666 | 9.750 | 9.666 | 9.750 |
| k=2 | 9.414 | 9.576 | 9.414 | 9.576 |
| k=3 | 9.167 | 9.400 | 9.167 | 9.400 |
| k=4 | 8.927 | 9.222 | 8.927 | 9.222 |
| k=5 | 8.689 | 9.037 | 8.689 | 9.037 |
| … | … | … | … | … |

This demonstrates the linkage between the conventional and the expanded gravity model: the traditional one yields the *instantaneous bilateral interaction between two cities* at a particular time, and the expanded one captures a whole spectrum of varying *unilateral interactions from one city to another over a span of time* across different time lags. From the expanded model, any spectrum density reflects the average interaction from this city to the other at a particular time lag. Table 3 summarizes the differences between them.

Table 3. Comparison between the conventional and the expanded gravity models

| Aspect | Conventional gravity model | Expanded gravity model |
|---|---|---|
| Origin | Conceptual analogue to the gravitational model in physics | Reasoning from geographic analysis |
| Usage | Description of spatial interaction | Analysis of spatio-temporal process of interaction |
| Objective | Instantaneous interaction | Interaction with a time lag and aggregation over time |
| Computation result | Gravity value | Gravity spectra |
| Data requirement | Cross-section data | Time-series data |
| Strengths | Simple computation | Revealing temporal process, containing the information captured by the conventional model |
| Weakness | Less informative | Requiring more data and complex computation |

## 4.3 Temporal Process of Spatial Interaction between Beijing and Tianjin

The Fourier analysis process of the expanded gravity model also provides interesting insights into the temporal process of spatial interaction or interdependency between Beijing and Tianjin.

In the second step of the gravity-model-based Fourier analysis, a cross power spectrum, *P(f)*, is obtained with equation (9). The estimated *P(f)* is a complex data series, represented as

$$\hat{P}(x,y) = \hat{p}(x,y) + \hat{q}(x,y)i.$$

The real part $\hat{p}(x,y)$ is referred to as *residual spectrum*, and the imaginary part $\hat{q}(x,y)$ is *quadrature spectrum*. The former reflects the energy distribution of spatial interaction between cities, and the latter can be used for analyzing periodicity in the spatio-temporal process.

In the case study, the relationship between the frequency *f* and the residual spectrum density, as shown in Figure 3(a), can be captured by a power function such as

$$p(f) = p_1 f^{-\beta} = 373.11 f^{-1.7704},$$

where the *spectral exponent* $\beta = 1.7704$, which relates to what is called *Hurst exponent* (Feder, 1988; Hurst *et al*, 1965; Mandelbrot, 1982). According to the relationship between the spectral exponent and the Hurst exponent such as (see e.g. Feder, 1988; Voss, 1985)

$$\beta = 2H + 1, \tag{20}$$

we have *Hurst exponent H*=0.3852 indicative of *fractional Brownian motion* (fBm, see Mandelbrot and Van Ness, 1968; Mandelbrot, 1982). Furthermore, the correlation coefficient between increments of interaction quantity is in the form (Peitgen and Saupe, 1988)

$$C = 2^{2H-1} - 1 = -0.1471 < 0.$$

This indicates a negative correlation of increments. If the interaction is increasing for some time $t_1$, then it tends to decrease for time $t_2 > t_1$. Obviously, the correlation coefficient value implies the *anti-persistence* fBm for interurban interactions, i.e., a growth at a time leads to a loss in the next stage, and *vice versa*. Anti-persistence may lead to erratical oscillation or even periodicity in interurban interactions, which will be examined by analysis of the quadrature spectrum.

Figure 3(b) shows the relationship between the frequency *f* and the quadrature spectrum density with an evident peak at $f = 0.03125$. This implies the presence of a definite *periodicity*. The corresponding period is around $T=1/f=32$ years. The periodicity underlies the population growth trends of the two cities, but is less evident and is revealed by the above spectral analysis. In fact, the selfsame periodicity has been revealed from the process of urbanization of China (Chen and Zhou, 2008). Maybe the spatio-temporal interaction of cities is dominated by the rules in dynamics of urbanization. Theoretical explanations for the periodicity in urban development can be also found in the Volterra-Lotka model of predator-prey ecological interaction (Dendrinos and Mullally, 1985), which will be expounded and illuminated in our future work.

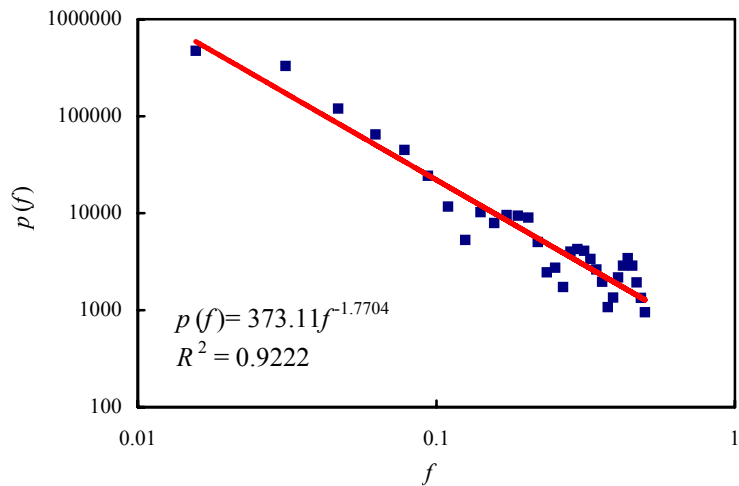

(a) Residual spectrum

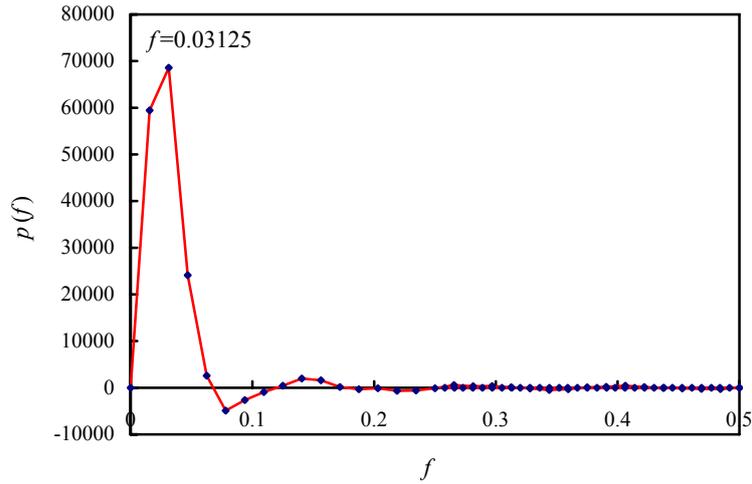

(b) Quadrature spectrum

**Figure 3. Cross spectrum analysis of spatial interaction between Beijing and Tianjin 1949-2000**

# 5 Conclusions

This paper develops an expanded gravity model for analyzing the complex spatio-temporal process in inter-urban attraction. By recognizing that spatial interaction takes time to develop, a time lag variable is introduced into the conventional gravity model. The new model measures the spatial interaction over a time span, and thus transforms the classical gravity model into time-series analysis. The core of the new model is a correlation function that is calibrated by Fourier analysis. The analysis yields a cross power spectrum, which can be decomposed to a residual spectrum and a quadrature spectrum. In our case study, the analysis of residual spectrum reveals *anti-persistence* for interurban interactions, i.e., a growth at a time leads to a loss in the next stage, and vice versa; and the analysis of quadrature spectrum reveals periodicity in the interactions.

The new model contains the information from the conventional model. In particular, when time lag is 0, the gravity value by the new model is simply the average of gravity values by the traditional model over the study period. With the time lag increases, the model reveals a declining spatial interaction that eventually becomes negligible after a period of time. The main benefits of the developed models are as follows. Firstly, it captures the essentials of spatial interaction in urban systems, and makes the notion of *urban gravity* to connect with the concept of *generalized* *energy*. Secondly, it presents a concept of urban *gravity spectrum* based on Fourier analysis, leading the pure spatial analysis based on interaction to spatio-temporal process analysis. Thirdly, it reveals asymmetry of spatial interactions between cities, which gives us a new insight into the spatial interaction of cities. The action from one city to another is not necessarily equal to the reaction in the reverse direction! Despite all the advantages, there are still some pending questions. One of the problems remaining to be solved next time, in a companion volume, is how to apply the new model to self-organized network of cities. After all, in the modern world of multi-centered society, spatial interaction can be observed among practically all pairs of major cities.

**Acknowledgements:** The first author would like to acknowledge the financial supports from the National Natural Science Foundation of China (Grant No. 40771061). The second author is grateful for the support from the National Natural Science Foundation of China (Grant No.40371035). Thanks are due to Professor Yixing Zhou at Peking University for his support.

## Appendix: Original data

**Table A. Population in Beijing and Tianjin 1949-2004 (in 10,000)**

| Year | Beijing | Tianjin | Year | Beijing | Tianjin |
|------|---------|---------|------|---------|---------|
| 1949 | 164.94 | 195.84 | 1977 | 452.84 | 347.19 |
| 1950 | 161.60 | 199.70 | 1978 | 467.00 | 358.45 |
| 1951 | 182.10 | 215.51 | 1979 | 495.21 | 380.57 |
| 1952 | 194.30 | 225.41 | 1980 | 510.40 | 392.62 |
| 1953 | 224.50 | 243.01 | 1981 | 522.60 | 400.75 |
| 1954 | 257.50 | 260.63 | 1982 | 534.00 | 410.70 |
| 1955 | 267.10 | 258.79 | 1983 | 547.10 | 419.57 |
| 1956 | 299.30 | 273.73 | 1984 | 558.10 | 437.09 |
| 1957 | 320.55 | 294.27 | 1985 | 572.50 | 445.69 |
| 1958 | 350.17 | 310.62 | 1986 | 586.80 | 447.05 |
| 1959 | 407.41 | 328.95 | 1987 | 601.00 | 454.68 |
| 1960 | 455.60 | 341.34 | 1988 | 614.20 | 467.38 |
| 1961 | 433.85 | 337.01 | 1989 | 630.60 | 476.62 |
| 1962 | 420.66 | 333.12 | 1990 | 640.10 | 485.44 |
| 1963 | 433.14 | 342.73 | 1991 | 648.40 | 490.32 |
| 1964 | 442.56 | 349.64 | 1992 | 656.30 | 494.80 |
| 1965 | 447.80 | 350.19 | 1993 | 668.70 | 499.98 |
| 1966 | 433.68 | 344.36 | 1994 | 683.80 | 504.03 |
| 1967 | 439.30 | 347.35 | 1995 | 696.90 | 507.94 |
| 1968 | 430.69 | 344.94 | 1996 | 709.70 | 513.15 |
| 1969 | 405.87 | 329.53 | 1997 | 722.70 | 515.36 |
| 1970 | 403.15 | 313.28 | 1998 | 733.70 | 521.37 |
| 1971 | 410.89 | 319.90 | 1999 | 747.20 | 528.68 |
| 1972 | 421.55 | 328.99 | 2000 | 760.70 | 532.51 |
| 1973 | 426.38 | 332.76 | 2001 | 780.10 | 535.22 |
| 1974 | 432.96 | 335.59 | 2002 | 806.90 | 541.14 |
| 1975 | 442.66 | 343.98 | 2003 | 830.80 | 549.74 |
| 1976 | 447.04 | 345.54 | 2004 | 854.70 | 556.17 |

**Sources**: (1) National Bureau of Statistics of China. 1999. *Comprehensive Statistical Data and Materials on 50 Years of New China*. Beijing: China Statistics Press. (2) Beijing Municipal Statistics Bureau. 2000-2005. *Beijing Statistical Yearbook*. Beijing: China Statistics Press. (3) Tianjin Municipal Statistics Bureau. 2000-2005. *Tianjin Statistical Yearbook*. Beijing: China Statistics Press. [in Chinese]